\documentclass[12pt]{article}
\pdfoutput=1

\usepackage{siunitx}

\newlength{\abstractwidth}
\usepackage{amsmath}
\usepackage{amsfonts}
\usepackage{amssymb}
\usepackage{latexsym}
\usepackage{braket}
\allowdisplaybreaks

\usepackage{epsf}
\usepackage{color}
\usepackage{graphicx}
\usepackage{tikz}
\usepackage{dsfont}
\usepackage{subfigure}

\usepackage{hyperref}
\definecolor{darkred}{rgb}{0.8,0.1,0.1}
\hypersetup{colorlinks=true, linkcolor=darkred, citecolor=blue, linktoc=page}

\usetikzlibrary{arrows,shapes,positioning}
\usetikzlibrary{decorations.markings}
\usepackage[rightcaption]{sidecap}
\tikzstyle arrowstyle=[scale=1]
\tikzstyle directed=[postaction={decorate,decoration={markings,
    mark=at position .65 with {\arrow[arrowstyle]{stealth}}}}]
\tikzstyle reverse directed=[postaction={decorate,decoration={markings,
    mark=at position .65 with {\arrowreversed[arrowstyle]{stealth};}}}]
\usetikzlibrary{positioning}

\renewcommand{\thefootnote}{\fnsymbol{footnote}}
\renewcommand{\thanks}[1]{\footnote{#1}}
\newcommand{\starttext}{
\setcounter{footnote}{0}
\renewcommand{\thefootnote}{\arabic{footnote}}}

\newcommand{\bea}{\begin{eqnarray}}
\newcommand{\eea}{\end{eqnarray}}
\newcommand{\be}{\begin{eqnarray}}
\newcommand{\ee}{\end{eqnarray}}

\def\){\right)}
\def\]{\right] }

\makeatletter\def\l@subsubsection#1#2{}%
\makeatother

\begin{document}
\starttext
\setcounter{footnote}{0}

\vskip 0.3in

\begin{center}

{\Large \bf Conformal Defect Solutions  in $N=2,D=4$ Gauged Supergravity}

\vskip 0.4in

{\large   Michael Gutperle and Matteo Vicino} 

\vskip .2in

 {\it Mani L. Bhaumik Institute for Theoretical Physics}\\
{ \it Department of Physics and Astronomy }\\
{\it University of California, Los Angeles, CA 90095, USA}\\[0.5cm]

 \vskip .2in
 
\begin{abstract}
 
 We consider conformal defect solutions in four dimensional  $N=2$ gauged supergravity. These solutions are constructed as a warped product of $AdS_2\times S^1$ over an interval with non-trivial electric and magnetic fields. We show for minimal gauged supergravity and for gauged supergravity with vector multiplets and abelian gauging that supersymmetric defect solutions are only possible when the geometry has a conical defect in either the bulk or the boundary metric.

\end{abstract}
\end{center}

\baselineskip=16pt
\setcounter{equation}{0}
\setcounter{footnote}{0}

\newpage

\section{Introduction}
\setcounter{equation}{0}
\label{sec1}

In many cases, conformal field theories contain not only local operators, but also extended objects or defects as well. Examples include Wilson lines \cite{Erickson:2000af,Drukker:2000rr}
 or surface defects  \cite{Gukov:2006jk,Gukov:2008sn} in $N=4$ SYM theories, surface defects in six-dimensional $(0,2)$ theories \cite{Ganor:1996nf}, and conformal defects in two dimensional CFTs \cite{Petkova:2000ip,Bachas:2001vj}. For conformal theories with holographic duals, the extended objects can often be described in the gravitational theory.  A special class of defects that preserve some (super)conformal symmetry on the  world-volume of the defect are called (super)conformal.  The so called Janus-ansatz \cite{Bak:2003jk} for a $p$-dimensional defect  in a higher dimensional spacetime contains an $AdS_{p+1}$ fiber whose isometry group realizes the conformal symmetry of the defect.  Additional symmetries, such as rotational invariance in the directions transverse to the defect or unbroken R-symmetries, can be realized by including spheres in the ansatz.
Unbroken superconformal symmetry implies the presence of Killing spinors or unbroken supersymmetries in the supergravity ansatz making the construction of exact solutions possible (see e.g.  \cite{DHoker:2007zhm,DHoker:2008lup,Bobev:2013yra,Chiodaroli:2009yw,Gutperle:2017nwo}). 

Four dimensional AdS gravitational theories with gauge fields and scalars have been used to model many strongly coupled three dimensional condensed matter systems including superfluids and superconductors, see e.g.   \cite{Hartnoll:2008vx,Balasubramanian:2008dm,Gubser:2008wv}. 
A simple model for  strongly coupled three dimensional CFTs is   four dimensional  AdS Einstein-Maxwell theory. The presence of a gauge fields allows the construction of charged defect solutions which  on the CFT side correspond to turning on a position dependent chemical potential for the charge dual to the gauge field. Such solutions were constructed in  \cite{Horowitz:2014gva} and for general forms of the chemical potential, it was found that the solutions break conformal invariance.  However, for a special choice of the gauge field there is the possibility to preserve a $SO(2,1)\times SO(2)$ subgroup of the three dimensional conformal group $SO(3,2)$ and such a solution is of the Janus type.

The goal of the present paper is to construct four dimensional conformal defect solutions in gauged $N=2$ supergravity.  In section \ref{sec2},  we embed the solutions \cite{Horowitz:2014gva} into minimal $N=2$ gauged supergravity  \cite{Freedman:1976aw} and generalize the solution to have both non-trivial electric and magnetic fields.   In section \ref{sec3}, we analyze the BPS conditions for the conformal defect solutions and show that for the conformal defect solutions there is a clash between supersymmetry and the regularity of the geometry.  We show that the solution breaks supersymmetry if we demand that there is no conical defect singularity present in the bulk metric and the boundary metric.  Four dimensional  gauged supergravity can be coupled to  vector  and hyper matter multiplets. In section \ref{sec4}, we consider coupling gauged $N=2$ supergravity to vector multiplets \cite{Andrianopoli:1996cm}. For abelian gaugings, we show that supersymmetry forces the scalars to be constant. In section \ref{sec5}, we discuss open questions and directions for future work.  We present our Clifford algebra conventions  and  basis for $AdS_2$ Killing spinors in Appendix \ref{appa}. In Appendix \ref{appb}, we present some details for the calculations in section \ref{sec3}.  In Appendix \ref{appc}, we prove that a more general ansatz for a conformal defect starting with an $AdS_2$ factor warped over a two dimensional Riemann surface $\Sigma$ with boundary reduces to the ansatz used above, i.e. supersymmetry implies the presence of an additional $U(1)$ isometry and hence the spacetime reduces to $AdS_2\times S^1$ warped over one spatial coordinate. This result is in line with classification theorems    found in \cite{Kunduri:2008rs,Cacciatori:2008ek}.

\section{Dyonic Conformal Defect Solution}
\setcounter{equation}{0}
\label{sec2}
The action for Einstein-Maxwell theory with negative cosmological constant  is given by
\be\label{einstein-maxwell}
S= \int d^4 x \sqrt{-g} \Big( {1\over 4} R - {1\over 4} F_{\mu\nu}F^{\mu\nu} + {3\over 2} {1\over L^2}\Big)
\ee
with the equations of motion take the following form
	\begin{align}\label{eq:EMGravity2}
	R_{\mu \nu}- 2 ({F_{\mu}}^{\rho}F_{\nu \rho} - \tfrac{1}{4} g_{\mu \nu} F_{\rho \sigma} F^{\rho \sigma}) + \frac{3}{L^2} g_{\mu \nu} &=0  \nonumber \\
	\partial_{\mu}(\sqrt{-g} F^{\mu \nu} )&= 0
	\end{align}
The conformal defect solution in the boundary CFT exhibits an $SO(2,1)\times SO(2)$ isometry  which can be realized by taking an $AdS_2\times S^1$ space warped over a radial coordinate.  We construct the most general solution which is asymptotically $AdS_4$ and has  nonzero  electric and magnetic components in the field strength $F_{\mu\nu}$.

%
\begin{align}\label{eq:spacetime}
	 ds^2 &= \frac{L^2}{\lambda^2 (1-y^2)^2} \bigg[ \rho^2\frac{-dt^2 + d\eta^2}{\eta^2} + y^2 f(y) d\phi^2 + \frac{4 \lambda^2 dy^2}{f(y)}\bigg]  \\
	A &= \frac{L \alpha } {\eta}  dt+ L \lambda \beta y^2  d\phi \\
	f(y) & = 3 - 3 y^2 + y^4 - 
 {\lambda^2 \over \rho^2}(1 - y^2)^2 + {\lambda^4\over \rho^4} (\alpha^2 + \beta^2 \rho^4) (1 - y^2)^3	\end{align}
  This solution is that of the analytically continued Reissner-Nordstr\"om black hole \cite{Kunduri:2008rs}
	\begin{equation}
	\begin{split}
	& ds^{2} = \psi^{2} \left( A_{0} r^2 dv^2 + 2 dv dr \right) + \frac{d\psi^2}{P(\psi)} + P(\psi) d\phi^2 \\
	& F = a \; dr \wedge dv + b \psi^{-2} d\psi \wedge d\phi
	\end{split}
	\end{equation}
with $P(\psi) = A_{0} - c(2\psi)^{-1} - (a^2+b^2) \psi^{-2} - \Lambda \psi^{2}/3$ under the identifications
	\begin{equation}
	\begin{split}
	& A_{0} = -\rho^{-2} \\
	& c = -\frac{2 \left(\lambda ^2-\rho ^2+ \lambda ^4(\alpha^2 \rho^{-2}+\beta ^2\rho ^2)\right)}{\lambda ^3 \rho ^2} \\
	& a = \alpha \rho^{-2} \\
	& b = \beta
	\end{split}
	\end{equation}
for $\psi^{-1} = \lambda(1-y^2)$ and $L=1$. 
 The condition that at $y=0$ the space closes off smoothly without a conical deficit, imposes a condition on the four parameters $\alpha,\beta,\rho$ and $ \lambda$.
  \be \label{eq:conical_constraint}
\frac{\lambda^4}{\rho^4} \left(\alpha^2 + \beta^2 \rho^4 \right) = 2 \lambda + \frac{\lambda^2}{\rho^2} - 3
 \ee
 With this condition imposed the function $f$ only depends on $\rho$ and $\lambda$
 \be
 f(y)=  3 - 3 y^2 + y^4 - 
 {\lambda^2 \over \rho^2}(1 - y^2)^2+ \left(2\lambda +{\lambda^2\over \rho^2}-3\right) (1 - y^2)^3	
 \ee
 In general, there is  an allowed parameter range for $\rho$ and $\lambda$ outside of which the function $f(y)$ develops a zero in the range $ [0,1]$ and the solution becomes singular.  For example, the allowed parameter range for $\lambda$ is given by $1\leq \lambda \lesssim 4.43$ for $\rho^2 =1 $.  The choice $\lambda=\rho=1$ corresponds to the $AdS_4$ vacuum with vanishing electromagnetic fields in an $AdS_2\times S^1$ slicing. As we shall see in the next section, solutions with $\rho^2\neq 1$ correspond to boundary spaces with a conical defect.

\subsection{Holography}

The conformal boundary of the metric (\ref{eq:spacetime}) is located at $y = 1$, but the metric is not in Fefferman-Graham (FG) form. However, it is straightforward to construct a FG coordinate $z$ near the boundary as a power series solution to
\be
 {dz\over z}=  {-2 d y\over (1-y^2) \sqrt{f(y)}}
\ee
This equation can be solved perturbatively to yield
\be
y = 1-z -{z^2\over 2} + {2 \lambda^2-\rho^2\over 2\rho^2}z^3-{32 \lambda^4(\alpha^2 +\beta^2 \rho^4)  + 8 \lambda^2 \rho^2 - 17 \rho^4 \over 24 \rho^4}z^4+ o(z^5)
\ee
and the metric (\ref{eq:spacetime}) becomes  

	\begin{align}
	ds^2 &= L^2  {dz^2 \over z^2} +  {L^2\over z^2} \Big(   g^{(0)}_{ij}+ z^2 g^{(2)}_{ij}+ z^3 g^{(3)}_{ij}  \Big) dx^i dx^j+ o(z^4)
	\end{align}
	with 
	\bea \label{eq:boundary_metric}
	 g^{(0)}_{ij} dx^i dx^j&=&  {1\over 4\lambda^2} \left( \rho^2{-dt^2+d\eta^2 \over \eta^2} + d\phi^2\right)\nonumber\\
	 g^{(2)}_{ij} dx^i dx^j&=& {1\over 2 }  \left( {-dt^2+d\eta^2 \over \eta^2} - {d\phi^2 \over \rho^2}\right)\ \nonumber\\
	 g^{(3)}_{ij} dx^i dx^j&=&  {2( \lambda^4 (\alpha^2 + \beta^2 \rho^4) + \lambda^2 \rho^2 - \rho^4) \over 3 \lambda^2 \rho^2} \Big(  {dt^2 -d\eta^2 \over \eta^2} + {2\over \rho^2} d\phi^2\Big)
	\eea
Following the standard holographic dictionary, $g^{(0)}$ is the metric of the $AdS_2\times S^1$ boundary, and $g^{(2)}$ is determined by $g^{(0)}$. Note that  for $\rho^2 \neq 1$ the boundary is conformal to $\mathbb{R}^{1,2}$ with a conical defect at $\eta=0$. The next term in the FG expansion $g^{(3)}$ determines the expectation values of the  stress tensor. Since we have an odd dimensional boundary, there is no conformal anomaly and
\be
\langle T_{ij}\rangle = {3\over 16 \pi G_N} g^{(3)}_{ij}
\ee
Note that the stress tensor is indeed traceless in agreement with conformal symmetry. The near boundary behavior of the gauge field in FG coordinates is given by
\be\label{fggauge}
A= \frac{L \alpha } {\eta}  dt+ L \lambda \beta (1 - 2 z) d\phi + o(z^2)
\ee
The standard holographic dictionary for a gauge fields identifies the $z^0$ term as a source and the $z^{1}$ term as an expectation value of the dual current $j_{\mu}$ in the CFT.  As discussed in \cite{Horowitz:2014gva}, the first term in (\ref{fggauge}) can be interpreted as  a chemical potential for the current. After a conformal transformation from $AdS_2\times S^1$ to $\mathbb{R}^{1,2}$, it takes the form $\mu(r)= L a_{\lambda}/r$.  This corresponds to a point charge defect localized at the origin $r=0$. The second term in (\ref{fggauge})  can be interpreted as a source and expectation value of $j^\phi$  \cite{Albash:2009iq,Montull:2009fe,Keranen:2009re,Dias:2013bwa}. As in \cite{Horowitz:2014gva}, the entanglement entropy of the defect can be analyzed using the Ryu-Takayanagi prescription \cite{Ryu:2006bv}. Extremal surfaces centered on the defect at $\eta = 0$ are given by $\eta = \eta_{0}$. The entanglement entropy of a region defined by $\eta \leq \eta_{0}$ is given by
	\begin{equation}
	S = \frac{1}{4G_{N}} \int d\phi dy \sqrt{g_{\phi \phi} g_{yy} } = \frac{\pi L^{2}}{ \lambda G_{N}} \int_{0}^{y_{\Lambda}} dy  \frac{y}{(1-y^{2})^{2}}
	\end{equation}
with $y_{\Lambda} \sim 1$ a UV cutoff. Note, the entanglement entropy $S$ does not depend on $\eta$. We can study the entanglement entropy of the defect alone by considering the quantity \newline 
	\begin{equation}
	\Delta S = S(\lambda) - S(1)
	\end{equation}
 Matching the circumference of circles near the asymptotic boundary requires
$ \lambda (1-\tilde{y}^{2}_{\Lambda}) = 1- y^{2}_{\Lambda} $ and leads to a defect entanglement entropy of
	\begin{equation}
	\Delta S = \frac{\pi L^{2}}{2 G_{N}} \left(1-\frac{1}{ \lambda } \right)
	\end{equation}	
	
\section{ Minimal Gauged Supergravity }
\label{sec3}

The field content of minimal $N=2, D=4$ gauged supergravity consists of a graviton $g_{\mu \nu}$, a pair of Majorana gravitini $(\psi_{\mu}^{1}, \psi_{\mu}^{2})$, and a photon $A_{\mu}$. The Einstein-Maxwell action (\ref{einstein-maxwell}) is the action of the purely bosonic sector. The full action includes the additional fermionic terms
	\begin{equation} \label{eq:minimal_sugra_action}
	\frac{1}{\sqrt{-g}}(\mathcal{L}_{\psi} + \mathcal{L}_{int}) = -\frac{1}{2} \bar{\psi}_{\mu} \gamma^{\mu \nu \rho} \mathcal{D}_{\nu} \psi_{\rho} - \frac{i}{8} \left(F + \hat{F} \right)^{\mu \nu} \bar{\psi}_{\rho} \gamma_{ [\mu} \gamma^{\rho \sigma} \gamma_{\nu]} \psi_{\sigma} + \frac{1}{2L} \bar{\psi}_{\mu} \gamma^{\mu \nu} \psi_{\nu}
	\end{equation}
with gauge covariant derivative $\mathcal{D}_{\mu} = \partial_{\mu} + \frac{1}{4} \omega_{\mu}^{ab} \gamma_{ab} - \frac{i}{L} A_{\mu}$. In the above, the gravitini have been combined into a Dirac spinor $\psi_{\mu} = \psi_{\mu}^{1} + i \psi_{\mu}^{2}$ with charge $e=1/L$ and $\hat{F}_{\mu \nu}$ is defined by   $\hat{F}_{\mu \nu}= F_{\mu \nu} - \text{Im}(\bar{\psi}_{\mu} \psi_{\nu})$.  For classical solutions, $\psi_{\mu} = 0$ and  the condition for unbroken supersymmetry of the background is the vanishing of the gravitino supersymmetry variation
\be\label{eq:gauged-killing-spinor}
& \delta \psi_{\mu} =\hat{\nabla}_{\mu} \epsilon = \left( \partial_{\mu} + \frac{1}{4} {\omega_{\mu}}^{ab} \gamma_{ab}- \frac{i}{L} 	A_{\mu} + \frac{1}{2L} \gamma_{\mu} +\frac{i}{4}{F}_{ab}\gamma^{ab} \gamma_{\mu} \right)  	\epsilon = 0
\ee
Given the AdS$_{2}$ isometry, we decompose the Killing spinors as a tensor product
	\begin{equation}
	\epsilon = \sum_{k, \kappa} \psi_{k}^{(\kappa)}(t,\eta) \otimes \chi_{k}^{(\kappa)}(y,\phi)
	\end{equation}
where $ \psi_{k}^{(\kappa)}$ satisfies the AdS$_{2}$ Killing spinor equation
	\begin{equation}
	\left( \partial_{\hat{\mu}} + \frac{1}{4} \hat{\omega}_{\hat{\mu}}^{\hat{a} \hat{b}} \tilde{\gamma}		_{\hat{a} \hat{b} } +  \frac{\kappa}{2} \tilde{\gamma}_{\hat{\mu}} \right) \psi^{(\kappa)}_{k} =0
	\end{equation}
with hatted indices denoting AdS$_{2}$ directions. Integrability requires $\kappa^{2} = 1$ and for each $\kappa$ there are two linearly independent solutions labelled by $k$. A Clifford algebra basis adapted to this decomposition and the explicit form of $\psi_{k}^{\pm}$ in this basis can be found in Appendix \ref{appa}. Multiplying the AdS$_{2}$ Killing spinor equation by the chirality matrix $\tilde{\gamma}_{\ast}$ shows we can choose $\psi_{k}^{\pm}$ to satisfy $ \tilde{\gamma}_{\ast} \psi_{k}^{\pm} = \psi_{k}^{\mp} $ which will be needed in the reduction. The reduction gives identical and decoupled equations for $k=1,2$ so any solution is automatically 1/2 BPS. In what follows, we drop the subscript $k$.

Applying the reduction to the following combination of BPS variations
	\begin{equation}
	\delta \psi_{t} - \gamma_{01} \delta \psi_{\eta} = - \frac{i \alpha}{\eta} \epsilon
	\end{equation}
shows that $\alpha = 0$ and only a purely magnetic solution can be supersymmetric. The AdS$_{2}$ components of the BPS equation (\ref{eq:gauged-killing-spinor}) gives
	\begin{equation}
	\left[ \left( \frac{1}{2} \mp \frac{\lambda(1-y^2)}{2\rho} \right) - \frac{\lambda(1-y^2)^2}{4Ly} A_{\phi}' \sigma_{2} \right] \chi^{\pm} = -\frac{\sqrt{f}}{4\rho}(1-y^2)^2 \frac{d}{dy} \left( \frac{\rho}{1-y^2} \right) \sigma_{3} \chi^{\mp} 
		\end{equation}
which requires $\chi^{\pm} \propto \ket{\pm}_{y} $ modulo the discrete symmetry $\lambda \rightarrow -\lambda$ and $ \chi^{\pm} \rightarrow \sigma_{3} \chi^{\pm}$. We therefore choose $\chi^{\pm} = h_{\pm}(y) e^{in\phi/2} \ket{\pm}_{y}$ with $n \in \mathbb{Z}$. The $\phi$ and $y$ components give the algebraic and differential equations
	\begin{equation}
	\begin{split}
	& \left( \frac{n}{2} -\frac{1}{L} A_{\phi} \mp \frac{\sqrt{f}(1-y^2)}{4\lambda} \frac{d}{dy}\left( \frac{y\sqrt{f}}{1-y^2} \right) \right) h_{\pm} = \left( -\frac{\sqrt{f}(1-y^2)}{4L} A_{\phi}' \pm \frac{y\sqrt{f}}{2\lambda(1-y^2)} \right) h_{\mp}
 \\
	& \frac{dh_{\pm}}{dy} = \left(- \frac{1}{\sqrt{f}(1-y^2)} \pm \frac{\lambda(1-y^2)}{2Ly\sqrt{f}} A_{\phi}' \right) h_{\mp}
	\end{split}
	\end{equation}
These equations are solved by
	\begin{equation}\label{eq:yphieq1}
	\begin{split}
	& h_{+}^2 + h_{-}^2 = \frac{\rho}{1-y^2} \\
	&h_{+}^2 - h_{-}^2 = \lambda+\frac{1}{2}\rho \lambda \left(n-\rho^{-1} \right)(1-y^2)
	\end{split}
	\end{equation}
subject to the conditions
	\begin{equation} \label{eq:susy_constraint}
	\begin{split}
	 & \lambda = 2(n+\rho^{-1})^{-1} \\
	 & 4\beta = n^2-\rho^{-2}
	 \end{split}
	 \end{equation}
Thus we have a family of supersymmetric solutions labelled by the parameters $n, \rho$. The conditions of no conical defect in the bulk (\ref{eq:conical_constraint})
 \begin{equation*}
\lambda^4 \beta^2 = 2 \lambda + \frac{\lambda^2}{\rho^2} - 3
 \end{equation*}
and no conical defect in the boundary (\ref{eq:boundary_metric})
\begin{equation*}
\rho^2 =1
\end{equation*}
are mutually incompatible with the constraints from supersymmetry (\ref{eq:susy_constraint}) with the exception of vacuum AdS$_{4}$ with $\lambda = \rho = n = 1$ and $\beta = 0$. Thus a non-trivial supersymmetric solution must either have a conical defect in the bulk at $y=0$ or in the boundary at $\eta = 0$.

\section{Charged Supersymmetric Defects  with Vector \newline Multiplets }
\setcounter{equation}{0}
\label{sec4}
In this section, we generalize the construction of magnetically charged defect solutions in gauged supergravity to the case where vector multiplets are present.  We first present our conventions for $N=2, D=4$ gauged supergravity with vector multiplets following \cite{Cacciatori:2008ek}.

\subsection{$N=2, D=4$ Gauged Supergravity with Vector Multiplets}
In addition to the previous field content, we now have complex scalars $z^{\alpha}$, Dirac fermions $\lambda^{\alpha}$, and vectors $A_{\mu}^{\alpha}$ with $\alpha = 1, \dots , n_{\text{V}}$, where $n_{\text{V}}$ is the number of vector multiplets. Supergravity theories with this matter content can be specified using the techniques of special geometry. It is convenient to introduce a new index $I = 0, 1, \dots , n_{\text{V}}$ and label the vectors as $A_{\mu}^{I}$ with $I=0$ the graviphoton. The complex scalars $z^{\alpha}$ will parameterize a special K{\"a}hler manifold 
	\begin{equation}
	\mathcal{V} = \begin{pmatrix} X^{I} \\ F_{I} \end{pmatrix}
	\end{equation}
with covariantly holomorphic sections  $\mathcal{D}_{\bar{\alpha}} \mathcal{V} = \partial_{\bar{\alpha}} \mathcal{V}- \frac{1}{2} \left( \partial_{\bar{\alpha}} \mathcal{K} \right) \mathcal{V} =0$.

The redundancy from introducing an additional coordinate $X^{0}$ is eliminated using the symplectic constraint
	\begin{equation}
	\left< \mathcal{V}, \overline{\mathcal{V}} \right> = X^{I} \bar{F}_{I} - F_{I} \bar{X}^{I} = i
	\end{equation}
Using a symplectic transformation, the section of the symplectic bundle can be brought into the form 
	\begin{equation}
	\mathcal{V} = e^{\mathcal{K}(z,\bar{z})/2} \begin{pmatrix} Z^{I}(z) \\ \partial_{I} F(Z) \end{pmatrix} =  e^{\mathcal{K}(z,\bar{z})/2} v(z)
	\end{equation}
with $F(Z)$ a homogeneous function of degree two called the prepotential. These two equations imply
	\begin{equation}
	 e^{-\mathcal{K}(z,\bar{z})} = -i \left< v , \bar{v} \right>
	\end{equation}
The supergravity theory is fully specified by the prepotential $F(Z)$ and a choice of the gauging of the $SU(2)$ R-symmetry. We will choose the $U(1)$ Fayet-Iliopoulos gauging and the gravitino charges will be denoted by $g_{I}$. The bosonic action of the theory is
	\begin{equation} \label{eq:lagrangian}
	e^{-1} \mathcal{L}_{\text{B}} = \frac{1}{2} R + \frac{1}{4} (\text{Im } \mathcal{N})_{IJ} F^{I}_{\mu \nu} F^{J \mu \nu} - \frac{1}{8}  (\text{Re } \mathcal{N})_{IJ} e^{-1} \epsilon^{\mu \nu \rho \sigma} F^{I}_{\mu \nu} F^{J}_{\rho \sigma} - g_{\alpha \bar{\beta}} \partial_{\mu} z^{\alpha} \partial^{\mu} \bar{z}^{\bar{\beta}}  - V
	\end{equation}
with scalar potential
	\begin{equation}
	V = - 2 g_{I} g_{J} \left[ (\text{Im } \mathcal{N})^{-1 | IJ} + 8\bar{X}^{I} X^{J} \right]
	\end{equation}
The kinetic matrix $\mathcal{N}_{IJ}$ is defined by
	\begin{equation}
	F_{I} = \mathcal{N}_{I J} X^{J} \hspace{1cm} \mathcal{D}_{\bar{\alpha}} \bar{F}_{I} = \mathcal{N}	_{IJ}  \mathcal{D}_{\bar{\alpha}} \bar{X}^{J}
	\end{equation}
and the K{\"a}hler metric is given by the expression $g_{\alpha \bar{\beta}} = \partial_{\alpha} \partial_{\bar{\beta}} \mathcal{K}$. 
	

The supersymmetry variations of the gravitino $\psi_{\mu}$ are
	\begin{equation}\label{eq:vectorkillingspinor}
	\begin{split}
	& \delta \psi_\mu = \hat{\nabla}_{\mu} \epsilon = \Big [ \partial_{\mu} + \frac{1}{4} \omega_{\mu}^{ab} \gamma_{ab} + \frac{i}{2} Q_{\mu} \gamma_{5} + ig_{I} A_{\mu}^{I} + g_{I} \gamma_{\mu} \left( \text{Im}(X^{I}) + i \gamma_{5} \text{Re}(X^{I}) \right) \\
	& + \frac{i}{4} \gamma^{ab} \left( \text{Im}(F_{ab}^{-I}X^{J}) - i \gamma_{5}  \text{Re}(F_{ab}^{-I}X^{J}) \right) \left(\text{Im }\mathcal{N} \right)_{IJ} \gamma_{\mu} \Big ] \epsilon
	\end{split}
	\end{equation}
and  that of the gauginos $\lambda^{\alpha}$ are
	\begin{equation}
	\begin{split}
	& \delta \lambda^{\alpha} = \Big [ \frac{i}{2} e^{\mathcal{K}/2} (\text{Im } \mathcal{N})_{IJ}\gamma^{ab} \left( \text{Im}(F_{ab}^{-J} \mathcal{D}_{\bar{\beta}}\bar{Z}^{I} g^{\alpha \bar{\beta}}) - i \gamma_{5}   \text{Re}(F_{ab}^{-J} \mathcal{D}_{\bar{\beta}}\bar{Z}^{I} g^{\alpha \bar{\beta}}) \right) + \\
	& \gamma^{\mu} \partial_{\mu}\left( \text{Re}(z^{\alpha}) -i\gamma_{5} \text{Im}(z^{\alpha}) \right) +2e^{\mathcal{K}/2}  g_{I}  \left(\text{Im}(\mathcal{D}_{\bar{\beta}} \bar{Z}^{I}g^{\alpha \bar{\beta}})-i\gamma_{5}\text{Re}(\mathcal{D}_{\bar{\beta}} \bar{Z}^{I}g^{\alpha \bar{\beta}})\right) \Big ] \epsilon
	\end{split}
	\end{equation}
where $\epsilon$ is a Dirac spinor. The K{\"a}hler connection
	\begin{equation}
	Q_{\mu} = -\frac{i}{2}	\left( \partial_{\alpha} \mathcal{K} \partial_{\mu} z^{\alpha} - \partial_{\bar{\alpha}} \mathcal{K} \partial_{\mu} \bar{z}^{\bar{\alpha}} \right)
	\end{equation}
appears in the supersymmetry transformations. 

\subsection{Integrability Conditions}
After incorporating the additional matter content and maintaining the $SO(2,1) \times SO(2)$ symmetry, our ansatz takes the form
	\begin{equation}
	\begin{split}
	& ds^2 = \frac{L^2}{\lambda^{2} (1-y^{2})^{2} } \left[ \rho(y)^2 \left(\frac{-dt^2 + d\eta^2}{\eta^2}\right) + y^{2} f(y) d\phi^{2} + \frac{4 \lambda^{2}}{f(y)} dy^{2} \right] \\
	& A^{I} = A_{\phi}^{I}(y) \hspace{.1cm} d\phi \hspace{1cm} z^{\alpha} = z^{\alpha}(y)
	\end{split}
	\end{equation}
By studying the integrability conditions, we can show that $\rho^2$ is constant. The equations of motion will then imply that the scalars $z^{\alpha}$ are also constant. The gauge field equations of motion are solved by
	\begin{equation} \label{eq:gauge_eom}
	\left( \text{Im} \mathcal{N} \right)_{IJ} \frac{dA_{\phi}^{J}}{dy} = \tilde{q_{I}} \frac{y}{\lambda \rho^2}
	\end{equation}
Using this relation and explicitly computing the integrability equations $ [ \hat{\nabla}_{\mu},  \hat{\nabla}_{\nu} ]\epsilon = 0 $ gives projection conditions of the form
\begin{equation} \label{eq:projection}
\begin{split}
& \gamma_{01} [ \hat{\nabla}_{t},  \hat{\nabla}_{\eta} ] \epsilon = \left( A_{0} + A_{2} \gamma_{2} +A_{23} \gamma_{23} + A_{013} \gamma_{013} \right) \epsilon = 0\\
& \gamma_{02} [ \hat{\nabla}_{t},  \hat{\nabla}_{\phi} ] \epsilon = \left( B_{0} + B_{2} \gamma_{2} +B_{01} \gamma_{01}+B_{23} \gamma_{23} + B_{013} \gamma_{013} \right) \epsilon = 0\\
& \gamma_{03} [ \hat{\nabla}_{t},  \hat{\nabla}_{y} ] \epsilon = \left( C_{0} + C_{2} \gamma_{2} + C_{3} \gamma_{3}+C_{01} \gamma_{01}+C_{23} \gamma_{23} + C_{012} \gamma_{012} + C_{013} \gamma_{013} \right) \epsilon = 0 \\
& \gamma_{23} [ \hat{\nabla}_{\phi},  \hat{\nabla}_{y} ] \epsilon = \left( D_{0} + D_{2} \gamma_{2} + D_{3} \gamma_{3} +D_{01} \gamma_{01}+D_{23} \gamma_{23} + D_{012} \gamma_{012} + D_{013} \gamma_{013} \right) \epsilon = 0
\end{split}
\end{equation}
The explicit expressions for these coefficients can be found in the appendix. The integrability conditions arising from the combinations $(\eta, \phi)$ and $(\eta, y)$ are equivalent to those arising from  $(t, \phi)$ and $(t, y)$. Due to the AdS$_{2}$ isometry, our ansatz will automatically be 1/2 BPS if it is supersymmetric. This can be seen from the Killing spinor decomposition of a previous section. We can thus impose at most one projection condition.
The tensor structure that appears in all four projection conditions is the one in  the first line of (\ref{eq:projection}).

Taking a linear combination of the first two lines of (\ref{eq:projection}) gives
	\begin{equation}
	\left(\tilde{B}_{1} + \tilde{B}_{01} \gamma_{01} \right) \epsilon = 0
	\end{equation}
and in order for this to not impose any additional projections, we must require $\tilde{B}_{1} = \tilde{B}_{01} = 0$. $ \tilde{B}_{01} \sim \text{Re}(g_{I}X^{I}) \text{Im}(\tilde{q}_{I} X^{I}) -  \text{Re}(\tilde{q}_{I}X^{I}) \text{Im}(g_{I} X^{I}) = 0$ also implies $C_{01}=D_{01} =0$. Finally, taking a linear combination of the last two lines of (\ref{eq:projection}) gives a projection condition involving the additional tensor structure
	\begin{equation}
	\sim f \rho \rho' \left( \text{Im}(g_{I}X^{I}) \gamma_{3} + \text{Re}(g_{I}X^{I}) \gamma_{012} \right) \epsilon 
	\end{equation}
relative to the first line of (\ref{eq:projection}). By the same reasoning, this must vanish and in order to not impose any additional constraints on $\epsilon$, $\rho'=0$. The Einstein field equations are given by
	\begin{equation}
	E_{\mu \nu} = R_{\mu \nu} +(\text{Im } \mathcal{N})_{IJ} \left( F_{\mu}^{I \alpha} F_{\nu \alpha}^{J} -\frac{1}{4} g_{\mu\nu} F_{\alpha \beta}^{I} F^{J \alpha \beta} \right) -2 g_{\alpha \bar{\beta}} \partial_{\mu}z^{\alpha} \partial_{\nu} \bar{z}^{\bar{\beta}}-V g_{\mu \nu} = 0
	\end{equation}
Forming a linear combination to isolate the scalar kinetic terms gives
	\begin{equation}
	0 = \frac{4(1-y^2)^2 \lambda^2}{yf^2} E_{\phi \phi} - y(1-y^2)^2 f^2 E_{yy} = 2y(1-y^2)^2 f^2 g_{\alpha \bar{\beta}} \frac{dz}{dy}^{\alpha} \frac{d\bar{z}}{dy}^{\bar{\beta}} 
	\end{equation}
Since $g_{\alpha \bar{\beta}}$ is invertible, each term in the sum must vanish individually and all the scalars are constant.

\subsection{General Model Reduction}	
From the previous section, the scalars $z^{\alpha}$ are constant and by a field redefinition we can choose $z^{\alpha} = \bar{z}^{\alpha}$. Furthermore, we can choose a parameterization with Re$(X^{I})=0$. The gravitino variation becomes
	\begin{equation}\label{eq:vectorkillingspinor2}
	 \delta \psi_\mu =  \left( \partial_{\mu} + \frac{1}{4} \omega_{\mu}^{ab} \gamma_{ab} + ig_{I} A_{\mu}^{I} - i g_{I} X^{I}  \gamma_{\mu} - \frac{\lambda(1-y^{2})^{2}}{4L^2y} \mathcal{F} \gamma_{23}\gamma_{\mu} \right) \epsilon
	\end{equation}
where
	\begin{equation}
	\mathcal{F} = (\text{Im } \mathcal{N})_{IJ} \frac{dA_{\phi}^{I}}{dy} X^{J}
	\end{equation}
and the gaugino variation becomes
	\begin{equation}\label{eq:gauginovariation}
	\delta \lambda^{\alpha} = \left( \frac{\lambda(1-	y^{2})^{2}}{2L^2y} (\text{Im } \mathcal{N})_{IJ} \mathcal{D}_{\beta} X^{I} \frac{dA_{\phi}^{J}}{dy} g^{\alpha \beta} \gamma_{23} +  2 i g_{I}\mathcal{D}_{\beta} X^{I}g^{\alpha \beta} \right) \epsilon
	\end{equation}
In this case, the gravitino BPS equations become identical to that of the pure gauged supergravity if we make the identifications
	\begin{equation}
	-\frac{1}{L} A_{\phi} \leftrightarrow g_{I}A_{\phi}^{I}, \hspace{1cm}  \frac{1}{2L} \leftrightarrow -i g_{I}X^{I}, \hspace{1 cm} A'_{\phi} \leftrightarrow -i \mathcal{F}
	\end{equation}
The reduction of the gaugino variation (\ref{eq:gauginovariation}) leads to the pair of equations
	\begin{equation} \label{eq:gaugino_bps}
	g_{I} \mathcal{D}_{\beta} X^{I} = \tilde{q}_{I} \mathcal{D}_{\beta} X^{I} = 0
	\end{equation}
This relation will give us constraints between the gauge couplings $g_{I}$ and the ``charges" $\tilde{q}_{I}$ of the defect.  

\subsection{$F = -i Z^{0} Z^{1}$ } 
We now specialize to the model with prepotential
	\begin{equation}
 	F(Z^{I}) = -i Z^{0} Z^{1}
 	\end{equation}
and parameterization
	\begin{equation}
	Z^{0} =i \hspace{.6cm} Z^{1}=i \tau
	\end{equation}
The data of the supergravity theory can be calculated from the special geometry to obtain
	\begin{equation}
	\begin{split}
	& \mathcal{K} = - \log{\left( 2(\tau+\bar{\tau}) \right)} \\
	& 	\mathcal{N} = -i
	\begin{pmatrix}
	\tau & 0 \\
	0 & \frac{1}{\tau}
	\end{pmatrix} \\
	V = -\frac{4}{\tau + \bar{\tau}} & \left( g_{0}^{2} + 2 g_{0} g_{1} ( \tau + \bar{\tau} ) +g_{1}^{2} \tau 		\bar{\tau} \right) \\
	\end{split}
	\end{equation}
for the Kahler potential, kinetic matrix, and scalar potential respectively. For simplicity, we will assume $g_{0},g_{1}>0$. The scalar potential $V$ has an AdS vacuum with $\tau = \bar{\tau} = g_{0}/g_{1}$ and radius of curvature $L = 1/\sqrt{4 g_{0} g_{1}}$.
The gaugino BPS equations (\ref{eq:gaugino_bps}) imply
	\begin{equation}
	\tau = \frac{g_{0}}{g_{1}} = \frac{\tilde{q}_{0}}{\tilde{q}_{1}}
	\end{equation}
This condition is equivalent to the vanishing of $\tilde{B}_{01}$ from the integrability conditions. The expressions for $h_{\pm}$, $f$, and $\lambda$ are identical, while the gauge fields $A_{\phi}^{I}$ are given by
	\begin{equation}
	A_{\phi}^{I} =- \frac{1}{4g^{I}} (n-\rho^{-1}) y^2
	\end{equation}
Note that $g_{I} A_{\phi}^{I} = -\frac{1}{2}(n-\rho^{-1})y^2$ as in the minimal gauged supergravity modulo a minus sign, which is conventional.
\section{Discussion }

In the present  paper we investigated  the question of whether supersymmetric conformal defect solutions exist in four dimensional AdS gauged supergravity. We considered a simple ansatz where the four dimensional geometry is given by an $AdS_2\times S^1$ factor warped over a single coordinate with non-trivial electric and magnetic field components.  
For minimal gauged supergravity without additional vector multiplets, the most general solutions of the equations of motion are double analytic continuations of black hole solutions.  We showed that  no supersymmetric solutions other than AdS$_{4}$ exist if we demand the absence of a conical defect in both the bulk and  boundary metrics. If one adds vector multiplets and uses the $U(1)$ Fayet-Iliopoulos gauging, we showed that demanding supersymmetry sets the scalars to be constant and the same conclusions as in the case without additional vector multiplets are reached. 
It is an interesting question whether a more generalized setup allows for supersymmetric solutions  with non-singular geometry.  Two generalizations which come to mind are  adding hypermultiplets and using non-abelian gaugings of isometries of  the vector and hypermultiplet scalar manifolds. We leave this question for future work.

\setcounter{equation}{0}
\label{sec5}

\section*{ Acknowledgements}

The work  of MG is supported in part by the National Science Foundation under grant PHY-16-19926.  MV is grateful to the Mani L.  Bhaumik Institute for Theoretical Physics for support.

\appendix

\section{Clifford Algebra Basis}
\label{appa}

A convenient basis for the $d=4$ Clifford algebra is given by
	\begin{equation}
	 \gamma_{0} = i \sigma_{2} \otimes 1 \hspace{.5cm}  \gamma_{1} = \sigma_{1} \otimes 1 \hspace{.5cm} \gamma_{2}=\sigma_{3} \otimes \sigma_{1} \hspace{0.5 cm} \gamma_{3} = \sigma_{3} \otimes \sigma_{3}
	\end{equation}
where $\tilde{\gamma}_{0}=i \sigma_{2}$ and $\tilde{\gamma}_{1}=\sigma_{1}$
form a basis of the $d=2$ Clifford algebra with chirality matrix $\tilde{\gamma}_{\ast} = \sigma_{3}$. In this basis, the Killing spinors of
	\begin{equation}
	ds^{2}_{AdS_{2}} = L^{2} \left( \frac{ -dt^{2} + d\eta^{2}}{\eta^{2}} \right) 
	\end{equation}
are given by
	\begin{equation}
	\psi_{1}^{\pm}=\frac{1}{\sqrt{\eta}} \begin{pmatrix}1 \\ \pm1 \end{pmatrix} \hspace{.5cm}
	\psi_{2}^{\pm}=\frac{1}{\sqrt{\eta}} \begin{pmatrix} t + \eta \\ \pm(t - \eta) \end{pmatrix}
	\end{equation}

\section{Explicit Expressions}
\label{appb}

The coefficients appearing in the expansion of the integrability conditions (\ref{eq:projection})  are given by

	\begin{align}
	&A_{0} = \frac{\left(1-y^2\right)^4 \left(\text{Im}(\tilde{q}_{I}X^{I})^2+\text{Re}(\tilde{q}_{I}X^{I})^2\right)+4 L^2 f^2 \left(4 L^2 \left(\text{Im}(g_{I}X^{I})^2+\text{Re}(g_{I}X^{I})^2\right)-y^2 \rho^2\right)}{8 L^2 \left(1-y^2\right)^2 \eta ^2 \lambda ^2 f}-\frac{1}{2 \eta ^2}\nonumber\\
	& A_{2} = -\frac{i y \text{Im}(\tilde{q}_{I}X^{I})\rho}{2 L \eta ^2 \lambda ^2} \nonumber\\
	& A_{23} = -\frac{i (\text{Im}(\tilde{q}_{I}X^{I}) \text{Im}(g_{I}X^{I})+\text{Re}(\tilde{q}_{I}X^{I}) \text{Re}(g_{I}X^{I}))}{\eta ^2 \lambda ^2} \nonumber\\
	& A_{013} = \frac{i y \text{Re}(\tilde{q}_{I}X^{I}) \rho}{2 L \eta ^2 \lambda ^2} \nonumber\\
	& B_{0} = \frac{y \sqrt{f} \rho \left(8 L^2 \left(\text{Im}(g_{I}X^{I})^2+\text{Re}(g_{I}X^{I})^2\right)+\rho \left(-\left(1+y^2\right) \rho-y \left(1-y^2\right) \rho' \right)\right)}{4 \left(1-y^2\right)^2 \eta  \lambda ^2} \nonumber\\
	& B_{2} = \frac{i y^2 \text{Im}(\tilde{q}_{I}X^{I}) \rho^2}{4 L \eta  \lambda ^2 \sqrt{f}}\nonumber \\
	& B_{01} = \frac{i y (-\text{Im}(g_{I}X^{I}) \text{Re}(\tilde{q}_{I}X^{I})+\text{Im}(\tilde{q}_{I}X^{I}) \text{Re}(g_{I}X^{I})) \rho}{2 \eta  \lambda ^2 \sqrt{f}}\nonumber \\
	& B_{23} = \frac{i y (\text{Im}(\tilde{q}_{I}X^{I}) \text{Im}(g_{I}X^{I})+\text{Re}(\tilde{q}_{I}X^{I}) \text{Re}(g_{I}X^{I})) \rho}{2 \eta  \lambda ^2 \sqrt{f}} \nonumber\\
	& B_{013} = -\frac{i y^2 \text{Re}(\tilde{q}_{I}X^{I}) \rho^2}{4 L \eta  \lambda ^2 \sqrt{f}}\nonumber\\
	& C_{0} = \frac{-y \left(1-y^2\right) \rho^2 \sqrt{f}'+\sqrt{f} \left(8 L^2 \left(\text{Im}(g_{I}X^{I})^2+\text{Re}(g_{I}X^{I})^2\right)+\rho \left(-\left(1+y^2\right) \rho-y \left(1-y^2\right) \rho' \right)\right)}{2 \left(1-y^2\right)^2 \eta  \lambda  \rho} \nonumber\\
	& C_{2} = \frac{i \left(\left(1-y^2\right) \text{Im}(\tilde{q}_{I}X^{I}) \sqrt{f}'+\sqrt{f} \left(2 y \text{Im}(\tilde{q}_{I}X^{I})-\left(1-y^2\right) \left(Q_{y} \text{Re}(\tilde{q}_{I}X^{I})+\text{Im}(\tilde{q}_{I}X^{I})'\right)\right)\right)}{4 L \eta  \lambda  f} \nonumber\\
	& C_{3} = -\frac{L \left(\text{Im}(g_{I}X^{I}) \sqrt{f}'+\sqrt{f} \left(Q_{y} \text{Re}(g_{I}X^{I})+\text{Im}(g_{I}X^{I})' \right)\right)}{\left(1-y^2\right) \eta  \lambda } \nonumber\\
	& C_{01} = \frac{i (-\text{Im}(g_{I}X^{I}) \text{Re}(\tilde{q}_{I}X^{I})+\text{Im}(\tilde{q}_{I}X^{I}) \text{Re}(g_{I}X^{I}))}{\eta  \lambda  \sqrt{f} \rho}\nonumber \\
	& C_{23} = \frac{i (\text{Im}(\tilde{q}_{I}X^{I}) \text{Im}(g_{I}X^{I})+\text{Re}(\tilde{q}_{I}X^{I}) \text{Re}(g_{I}X^{I}))}{\eta  \lambda  \sqrt{f} \rho}\nonumber \\
	& C_{012} = -\frac{L \left(\text{Re}(g_{I}X^{I}) \sqrt{f}'+\sqrt{f} \left(-\text{Im}(g_{I}X^{I}) Q_{y}+\text{Re}(g_{I}X^{I})' \right)\right)}{\left(1-y^2\right) \eta  \lambda } \nonumber\\
	& C_{013} = -\frac{i \left(\left(1-y^2\right) \text{Re}(\tilde{q}_{I}X^{I}) \sqrt{f}'+\sqrt{f} \left(\left(1-y^2\right) \text{Im}(\tilde{q}_{I}X^{I}) Q_{y}+2 y \text{Re}(\tilde{q}_{I}X^{I})-\left(1-y^2\right) \text{Re}(\tilde{q}_{I}X^{I})' \right)\right)}{4 L \eta  \lambda  f}  \nonumber\\
	&  D_{0} = \frac{1}{4 L^2 y \left(1-y^2\right) f} \big( y \left(1-y^2\right)^4 (\text{Im}(\tilde{q}_{I}X^{I})^2+ \text{Re}(\tilde{q}_{I}X^{I})^2)+L^2 f^2 \big(-16 L^2 y (\text{Im}(g_{I}X^{I})^2+ \text{Re}(g_{I}X^{I})^2)\nonumber \\  
	&\hphantom{D_{0} =} +4 y \rho^2+ 3 \rho \rho' -2 y^2 \rho \rho' - y^4 \rho \rho' +y \rho'^2-2 y^3 \rho '^2+y^5 \rho'^2+ y \left(1-y^2\right)^2 \rho \rho'' \big) \big) \nonumber\\
	& D_{2} = \frac{i y \rho \left(2 \left(1-y^2\right) \text{Im}(\tilde{q}_{I}X^{I}) \sqrt{f}'+\sqrt{f} \left(4 y \text{Im}(\tilde{q}_{I}X^{I})-\left(1-y^2\right) \left(Q_{y} \text{Re}(\tilde{q}_{I}X^{I})+\text{Im}(\tilde{q}_{I}X^{I})' \right)\right)\right)}{4 L \lambda \sqrt{f^3}}\nonumber \\
	& D_{3} = \frac{L y \rho \left(Q_{y} \text{Re}(g_{I}X^{I})+\text{Im}(g_{I}X^{I})' \right)}{\left(1-y^2\right) \lambda } \nonumber\\
	& D_{01} = \frac{2 i y (-\text{Im}(g_{I}X^{I}) \text{Re}(\tilde{q}_{I}X^{I})+ \text{Im}(\tilde{q}_{I}X^{I}) \text{Re}(g_{I}X^{I}))}{\lambda  f} \nonumber\\
	& D_{23} = -i g_{I}{A^{I}_{\phi}} '\nonumber \\
	& D_{012} = -\frac{L y \rho \left(\text{Im}(g_{I}X^{I}) Q_{y}-\text{Re}(g_{I}X^{I})' \right)}{\left(1-y^2\right) \lambda } \nonumber\\
	& D_{013} = -\frac{i y \rho \left(2 \left(1-y^2\right) \text{Re}(\tilde{q}_{I}X^{I}) \sqrt{f}'+\sqrt{f} \left(\left(1-y^2\right) (\text{Im}(\tilde{q}_{I}X^{I}) Q_{y} - \text{Re}(\tilde{q}_{I}X^{I})')+4 y \text{Re}(\tilde{q}_{I}X^{I})\right)\right)}{4 L \lambda  \sqrt{f^3}}
	\end{align}

\section{$AdS_2\times \Sigma_2$ Ansatz and the Emergence of an Additional Isometry}
\label{appc}

A more general ansatz incorporating a two dimensional Riemann surface $\Sigma_{2}$ is given by
	\begin{equation}\label{eq:Riemann}
	\begin{split}
	& ds^2 =L^2 \rho(z,\bar{z})^{2}\bigg( \frac{-dt^2 +d\eta^2}{\eta^2} \bigg) + L^2 f(z,\bar{z})^{2} 		\hspace{.05cm}dz d\bar{z} \\
	& A = A_{z}(z,\bar{z}) \hspace{.05cm}dz+ A_{\bar{z}}(z,\bar{z}) \hspace{.05cm} d\bar{z}
	\end{split}
	\end{equation}
The equations $\delta_{t} \psi_{\mu} = \delta_{\eta} \psi_{\mu} = 0$ give the projection conditions $\chi^{\pm} = h_{\pm}(z,\bar{z}) \ket{\pm}_{y} $ modulo a discrete symmetry. In what follows, we have dropped the subscript $k$ since identical equations hold for $k=1$ and $k=2$ and the solution will automatically be 1/2 BPS. As before, a component of the gauge field of the form $A_{t}(\eta)$ is inconsistent with the BPS equations. The equations $\delta_{z} \psi_{\mu} = \overline{\delta_{\bar{z}} \psi_{\mu}} = 0$ give
	\begin{equation}\label{eq:case1.1}
	\frac{\partial h_{+}}{\partial z} + \frac{1}{2f} \frac{\partial f}{\partial z} h_{+} -\frac{i}{L} A_{z} h_{+}-i  	f \left( \frac{1}{2} + \frac{i}{L \hspace{0.05cm} f^2} \left( \frac{\partial A_{z}}{\partial \bar{z}} - 		\frac{\partial A_{\bar{z}}}{\partial z} \right) \right) h_{-} 	=0
	 \end{equation}
	\begin{equation}\label{eq:case1.2}
	 \frac{\partial h_{-}}{\partial z} - \frac{1}{2f} \frac{\partial f}{\partial z} h_{-} -\frac{i}{L} A_{z} h_{-} = 0
	 \end{equation}
	 \begin{equation} \label{eq:case1.3}
	\frac{\partial {\bar{h}} _{+} }{\partial z} - \frac{1}{2 f} \frac{\partial f}{\partial z} {\bar{h}_{+}} + 		\frac{i}{L} A_{z} {\bar{h}_{+}}= 0 
	\end{equation}
	\begin{equation} \label{eq:case1.4}
	\frac{\partial {\bar{h}_{-}} }{\partial z} + \frac{1}{2f} \frac{\partial f}{\partial z} {\bar{h}_{-}} + 	\frac{i}	{L} A_{z} {\bar{h}_{-}} - i f \left( \frac{1}{2} - \frac{i}{L \hspace{0.05cm} f^2} \left( \frac{\partial 		A_{z}}{\partial \bar{z}} - \frac{\partial A_{\bar{z}}}{\partial z} \right) \right){\bar{h}_{+}}= 0
	\end{equation}
Equations (\ref{eq:case1.2}) and  (\ref{eq:case1.3}) can be solved immediately to yield
	\begin{equation}\label{eq:Az1}
	\frac{i}{L} A_{z} = \frac{\partial}{\partial z} \ln \left( \frac{h_{-}}{\sqrt{f}} \right) = \frac{\partial}		{\partial z} \ln \left( \frac{\sqrt{f}}{{\bar{h}_{+}}} \right)
	\end{equation}
This implies we must have
	\begin{equation} \label{eq:rho1}
	{\bar{h}_{+}} h_{-}=\bar{g}_{1}(\bar{z}) f
	\end{equation}
for some anti-holomorphic function $\overline{g}_{1}(\bar{z})$. Taking the linear combination ${\bar{h}_{+}} (\ref{eq:case1.1}) $+ ${\bar{h}_{-}} (\ref{eq:case1.2})$ + $h_{+} (\ref{eq:case1.3})$ + $h_{-} (\ref{eq:case1.4}) $ gives
	\begin{equation}
	g_{1}(z) \frac{\partial}{\partial z} \left( |h_{+}|^{2} + |h_{-}|^{2} \right) = i |h_{+}|^{2} |h_{-}|^{2}
	\end{equation}
where we have used (\ref{eq:rho1}) and $\bar{f}=f$. Taking the linear combination ${\bar{h}_{+}} (\ref{eq:case1.1}) $ - $\bar{h}_{-} (\ref{eq:case1.2})$  + $h_{+} (\ref{eq:case1.3})$ - $h_{-} (\ref{eq:case1.4}) $ gives
	\begin{equation}
	\frac{\partial}{\partial z} \left( |h_{+}|^{2} - |h_{-}|^{2} \right) = i \overline{g}_{1}	(\bar{z}) \frac{\partial}{\partial z} \frac{\partial}{\partial \bar{z}} \ln \left( \frac{ |h_{-}|^{2} }{ |h_{+}|^{2} } \right)
	\end{equation}
which can be immediately integrated in $z$. To obtain this equation, one needs to make use of both (\ref{eq:Az1}) and (\ref{eq:rho1}). Setting $u = |h_{+}|^{2} $, $v = |h_{-}|^{2}$, and then integrating and complex conjugating the previous equation gives
	\begin{equation}\label{eq:uv1}
	\begin{split}
	& g_{1}(z) \frac{\partial}{\partial z} \left( u + v \right) = i uv \\
	& u - v + i g_{1}(z) \frac{\partial}{\partial z} \ln \left( \frac{v}{u} \right) = \frac{d \tilde{g}_{2}(z)}{dz}
	\end{split}
	\end{equation}
for some holomorphic function $\tilde{g}_{2}(z)$. After performing the change of variables
	\begin{equation}
	u = e^{X+Y} \hspace{.5 cm} v = e^{X-Y}
	\end{equation}
and change of coordinates defined by $g_{1}(z) = \frac{d z}{d w}$, equation (\ref{eq:uv1}) becomes
	\begin{equation}\label{eq:uv2}
	\begin{split}
	& \frac{\partial}{\partial w} \left( 2 e^{X} \cosh{Y} \right) = i e^{2 X} \\
	& 2 e^{X} \sinh{Y} -2 i \frac{\partial Y}{\partial w} = \frac{d g_{2}(w)}{dw}
	\end{split}
	\end{equation}
Reality of $X$ and $Y$ requires
	\begin{equation}
	2i \frac{\partial Y}{\partial w} + \frac{d g_{2}}{dw} = -2i \frac{\partial Y}{\partial \overline{w}} + 		\frac{d \hspace{.04cm} \overline{g}_{2}}{d \overline{w}}
	\end{equation}
	which is solved by
	\begin{equation}
	Y = \frac{i}{2} \left( g_{2}(w) - \overline{g}_{2}(\overline{w}) \right) + \tilde{y}(w-\overline{w})
	\end{equation}
Solving for $e^{X}$ using the second line of (\ref{eq:uv2}) and substituting into the first gives
	\begin{equation}\label{eq:tilde1}
	0 = \left( \frac{\partial \tilde{y}}{\partial w} \right)^{2} + i \frac{d g_{2}}{d w} \frac{\partial \tilde{y}}		{\partial w} - \sinh{\left( i(g_{2}-	\overline{g}_{2})+2 \tilde{y} \right)} \frac{\partial^{2} \tilde{y}}		{\partial w^{2}}
	\end{equation}
Forming the combination (\ref{eq:tilde1}) - $\overline{(\ref{eq:tilde1})}$ gives
	\begin{equation}
	i \left( \frac{d g_{2}}{d w} - \frac{d \overline{g}_{2}}{d \overline{w}} \right)  \frac{\partial \tilde{y}}{\partial w} = 0
	\end{equation}
This equation is solved by $g_{2}(w) = a w + b$ with $a \in \mathbb{R}$. The case $ \frac{\partial \tilde{y}}{\partial w} = 0$ corresponds to the field strength vanishing identically. Thus, all the fields are only a function of $w-\bar{w} \propto y$. The Riemann ansatz simplifies to
	\begin{equation}
	\begin{split}
	& ds^2 =L^2 \rho(y)^{2}\bigg( \frac{-dt^2 +d\eta^2}{\eta^2} \bigg) + L^2 f(y)^{2} \hspace{.05cm}(d	\phi^2 + dy^2) \\
	& A = A(y) \hspace{.05cm}d\phi
	\end{split}
	\end{equation}
This ansatz is simple enough to use the Einstein-Maxwell equations directly. The $\phi$ component is the only non-trivial component of Maxwell's equation and is solved by
	\begin{equation}
	f(y) = c_{1}\rho(y)^{2} \frac{dA}{dy}
	\end{equation}
Next, the difference of the $\phi \hspace{.05cm} \phi$ and $y\hspace{.05cm}y$ components of Einstein's field equation is solved by
	\begin{equation}
	\frac{dA}{dy} = c_{2} \frac{2 \rho'(y)}{\rho(y)^{2}}
	\end{equation}
Thus, we only have 1 unknown function $\rho(y)$ and solving the BPS equations will recover the previous magnetic defect solution. Starting from an ansatz of AdS$_{2}$ warped over a Riemann surface $\Sigma_{2}$, we see that supersymmetry forces the existence of a $U(1)$ isometry $S^{1}$. In summary, we showed that a more general ansatz of  an $AdS_2$ factor warped over a Riemann surface reduces to $AdS_2\times S^1$ warped over a one dimensional interval.


\end{document}